# Collision of two action potentials in a single excitable cell

Christian FILLAFER[1], Anne PAEGER[1], Matthias F. SCHNEIDER[1],*

[1]Medical and Biological Physics
Faculty of Physics
Technical University Dortmund
Otto-Hahn-Str. 4
44227 Dortmund
Germany

*Address correspondence to: Matthias F. Schneider; Physics - Medical and Biological Physics
Technical University Dortmund; Otto-Hahn-Str. 4; 44227 Dortmund
(Germany); tel: +49-231-755-4139
email: matthias-f.schneider@tu-dortmund.de

# Abstract

*Background.* It is a common incident in nature, that two waves or pulses run into each other head-on. The outcome of such an event is of special interest, because it allows conclusions about the underlying physical nature of the pulses. The present experimental study dealt with the head-on meeting of two action potentials (AP) in a single excitable plant cell (*Chara braunii* internode).

*Methods.* The membrane potential was monitored at the two extremal regions of an excitable cell. In control experiments, an AP was excited electrically at either end of the cell cylinder. Subsequently, stimuli were applied simultaneously at both ends of the cell in order to generate two APs that met each other head-on.

*Results.* When two action potentials propagated into each other, the pulses did not penetrate but annihilated (N=26 experiments in n=10 cells).

*Conclusions.* APs in excitable plant cells did not penetrate upon meeting head-on. In the classical electrical model, this behavior is specifically attributed to relaxation of ion channel proteins. From an acoustic point of view, annihilation can be viewed as a result of nonlinear material properties the entire system.

*General significance.* The present results suggest that APs in excitable animal and plant cells belong to a similar class of nonlinear phenomena. Intriguingly, other excitation waves in biology (intracellular waves, cortical spreading depression, etc.) also annihilate upon collision and are thus expected to follow the same underlying principles as the observed action potentials.

# Graphical abstract

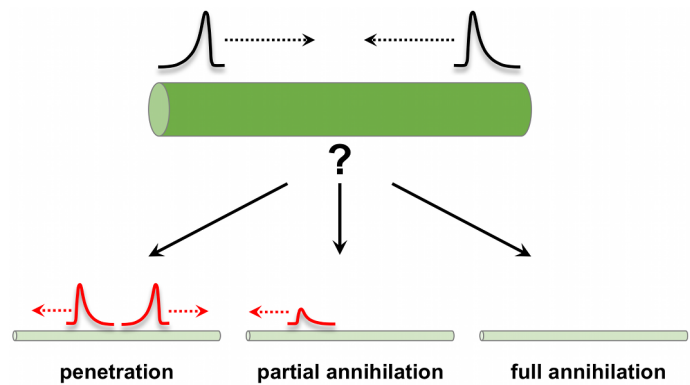

# Highlights

- When two pulses meet, they reveal information about their physical nature.
- Upon running into each other, two action potentials in an excitable plant cell annihilate
- Action potentials in plant cells and nerves are similar nonlinear phenomena.

## 1. Introduction

Action potentials (AP) are propagating all-or-none pulses that have been observed in excitable cells such as neurons, myocytes, plant cells (*e.g.* Charophytes), etc. These cells can assume different morphologies which commonly contain elongated cylindrical elements. APs can be stimulated at either end of such a cylinder, *i.e.* propagation in the antero- as well as retrograde direction is possible (in the neurophysiology literature the terms ortho- (in "natural" direction of fibre) and antidromic (against "natural" direction of fibre) are most often used). Thus, two pulses may run into each other. This leads to several interesting questions, for instance, do two APs superimpose and penetrate or do they partially/fully annihilate? The outcome of such an experiment is an important criterion, since it provides information about the physical nature of an action potential. Moreover, it has been suggested that a reduction of pulse collision events may drive the formation of directional transmission in nerve cell networks (K. Kaufmann, personal communication).

What is the present state of evidence? In neurophysiology, annihilation of two APs upon collision – due to the trailing refractory zone – is considered to be an established fact. This assertion provides the rationale for several experimental techniques. One of those is the "collision test", which is used, amongst others, to determine the level of natural (*i.e.* orthodromic) activity in a nerve trunk [1,2]. In a collision test, antidromic APs are stimulated artificially. Only those pulses which propagate in axons with no orthodromic activity will be detectable at the opposite end of the nerve, whereas the others will be extinguished by collision. Based on these ideas, a technique called "collision block" has been conceived [3]. The goal of the latter is to avoid overstimulation of an end organ as a result of high nerve activity. For this purpose, antidromic APs are stimulated artificially to eliminate orthodromic pulses. The results of collision test as well as collision block experiments provide evidence that APs annihilate upon meeting head on. However, since such studies assume the premise *that collision leads to annihilation* to be true they are not ideal to investigate it. Otherwise, to the best of our knowledge, only few systematic studies exist. In 1949, Tasaki demonstrated that two APs extinguish upon meeting head-on in single myelinated nerve fibres [4]. These results were corroborated in ventral nerve cord of earthworm (*Lumbricus hortensis*) [5], the stomatogastric nervous system of the crab (*Cancer borealis*) [5], as well as close to bifurcations in lobster axons [6]. In contradiction to these findings, Gonzalez-Perez et al. reported penetration of APs in ventral nerve cord from *Lumbricus terrestris* as well as *Homarus americanus* [7].

The potential importance of pulse collision for nervous network formation, the scarcity of research of the phenomenon and the possibility of different outcomes have motivated us to conduct the present study. So far, most experiments in the literature have been carried out with nerve trunks. The latter preparations are not ideally suited to obtain clear results. A nerve consists of several fibres with varying diameters which are usually excited and recorded by external electrodes. To be able to properly set up and interpret such experiments requires extensive experience. It has to be ascertained, for example, that the component fibre in which the pulses are excited is known. Otherwise, ambiguities remain (*e.g.* two different fibres may be stimulated and thus the pulses never come to meet each other which will be misinterpreted as penetration). In any case, it is difficult to judge from extracellular recordings from a nerve trunk whether annihilation in a component fibre is complete or partial. Thus, it seemed expedient to study the head-on meeting of two APs in a preparation that allows for unambiguous conclusions. Ideally, such a system is simple to set up so that the results can be easily reproduced/falsified by others. Giant internodes from *Charophytes* are well suited for this purpose. These cylindrical plant cells are one of the most extensively studied systems in excitable cell physiology [8–10]. Their large diameter (0.5-1 mm) and cell length (up to 15 cm) dramatically facilitate experimental work. Thus, we use *Chara braunii* internodes herein to study what happens when two APs run into each other.

## 2. Materials and Methods

*2.1 Materials.* All reagents used were purchased from Sigma-Aldrich (St. Louis, MO, USA) and were of analytical purity (≥99%).

*2.2 Cell cultivation and storage.* *Chara braunii* cells were cultivated in glass aquariums filled with a layer of soil (~2 cm), quartz sand (~1cm) and deionized water. The cells were grown under illumination from an aquarium light (14W, Flora Sun Max Plant Growth, Zoo Med Laboratories Inc., San Luis Obispo, CA, USA) at a 14:10 light:dark cycle at room temperature (~20°C). Prior to use, single internodal cells were stored for a minimum of 12 h in a solution containing 0.1 mM NaCl, 0.1 mM KCl and 0.1 mM $CaCl_2$ [8].

*2.3 Experimental setup.* A single internodal cell (6-12 cm long) was placed on a plexiglass chamber into which compartments (~5 x 5 x 10 mm; h x w x l) had been milled. Small extracellular sections (length ~5 mm) of the cell were electrically isolated against each other with vacuum grease (Dow Corning Corporation, Midland, MI, USA). The $K^+$-anesthesia technique [8] in combination with extracellular Ag/AgCl-wire electrodes was used for monitoring the cell membrane potential. The extracellular solutions contained 110 mM KCl in the outmost compartments and artificial pond water (APW) in all other compartments (1 mM KCl, 1 mM $CaCl_2$, 5 mM TRIS and 190 mM D-sorbitol; pH set to 7.0 with HCl). The potential between the virtual intracellular electrodes (KCl-compartments) and extracellular electrodes was recorded with voltage sensors (PS-2132; 100Hz sample rate; PASCO scientific, Roseville, CA, USA). All experiments were conducted at room temperature (20±2 °C).

*2.4 Controls and collision experiment.* Upon equilibration of the cell in the measurement chamber, an action potential was excited at either end of the cell by application of a rectangular current pulse across the outer compartments (Fig. 1A; 9V applied across a 1 MΩ resistor; stimulus duration: 50-100 msec; stimulus frequency: ~1AP per 15 min). This procedure ascertained that the internode was fully excitable and that the triggered AP propagated along the entire cell length. In order to generate a collision event, stimuli were applied simultaneously at both ends of the cell. In total, N=26 experiments were performed in which two APs met each other head-on (n=10 cells). All electrophysiological data for control and collision experiments are publicly available [11].

## 3. Results and Discussion

*3.1 Control experiments.* An action potential was excited in a *Chara* cell by application of a superthreshold electrical shock. Subsequently, the pulse propagated along the cell cylinder. Upon stimulation at the left end of the cell, the AP passed by the two potential sensors as expected from their geometrical setup (Fig. 1B and 2B). It was experimentally confirmed that stimulation at the right end of the cell led to propagation along the reversed path (Fig. 1C and 2C). These control experiments ascertained that both ends of the cell are electrically excitable and they established the characteristic appearance of an AP at the sensor locations. Typically, the duration of an action potential was in the range of ~10 s and the conduction velocity was ~5 mm $s^{-1}$ from which it is concluded that the pulse has a spatial extension or wavelength of ~5 cm. The pulse amplitude, duration as well as propagation velocity found herein are in accordance with the ones reported in the literature [8,9]. The pulse shape in Characean cells can be subject to slight variations (*e.g.* along the cell cylinder or with duration of an experiment). This is usually more pronounced in some cells than in others. Thus, additional control experiments were intermittently performed in course of an experiment in order to identify changes in pulse shape *that are not due to pulse collision*. These experiments revealed that broadening of an AP or small distortions can occur in absence of collision (see examples in [11]). Therefore, caution must be taken in order not to erroneously attribute such changes in pulse shape during a collision experiment to pulse penetration, etc.

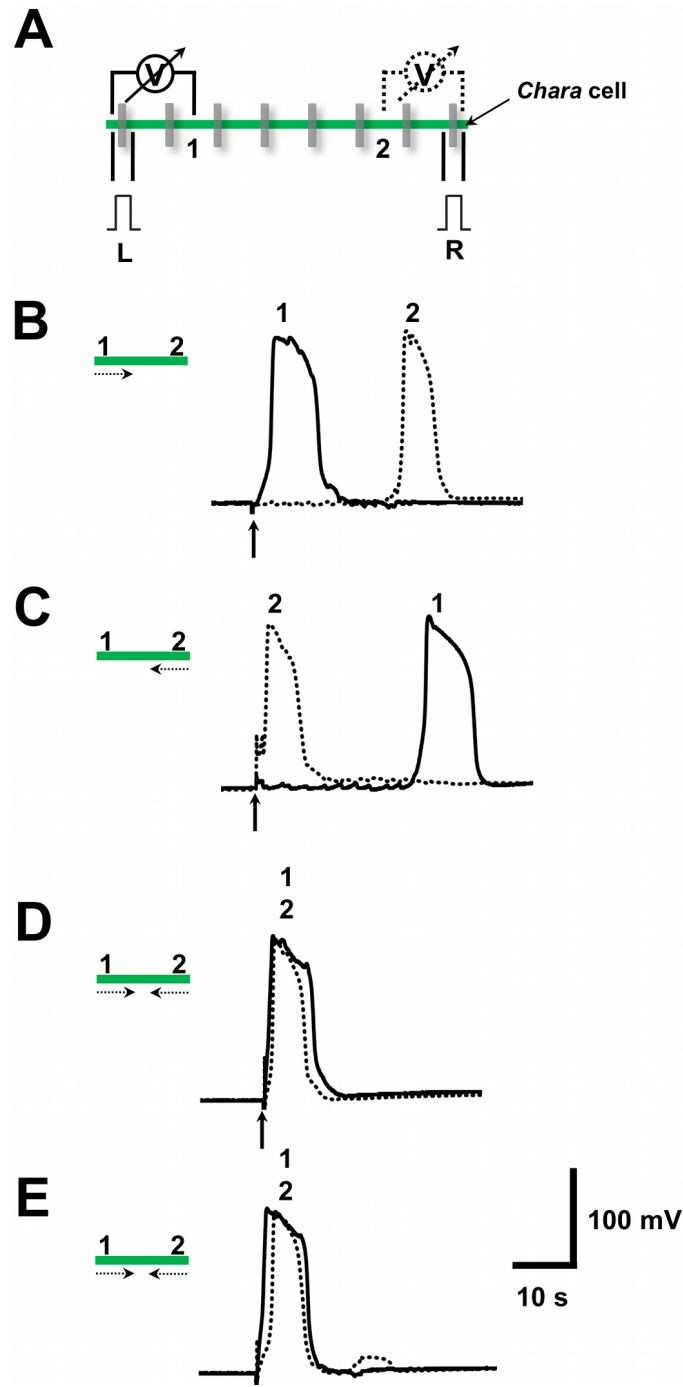

**Figure 1.** *Collision of action potentials (AP) in Chara braunii cells.* **(A)** The membrane potential was monitored at the two extremal regions (1) (solid line) and (2) (dotted line) of a *Chara* internode. By application of a superthreshold electrical stimulus, an AP was excited at the left (L) and right end (R) of the cell cylinder. **(B)** A single excitation at L led to the opposite propagation path as compared to a single excitation at R **(C)**. **(D, E)** Simultaneous excitation at L and R led to the head-on meeting of two APs in the central region of the cell. At each sensor only one AP was recorded which indicates that the pulses did not penetrate.

***3.2 Collision of two action potentials.*** When two APs are excited simultaneously at both ends of a cylindrical cell, they propagate towards the center and meet head-on. Prior to a discussion of the results, it is instructive to consider the possible outcomes of such an experiment. If the two APs superimpose and penetrate or reflect off of each other, one expects the appearance of two pulses at each sensor. The two pulses could manifest either as a fused event (the membrane potential deflection may assume, for example, an "M"-like shape) or as discrete peaks. Based on the spatial extension of an AP in *Chara* internodes as calculated above (~5 cm), two discrete pulse events should be observable at the extremal sensors if the pulses simply penetrate and if the cell length is larger than ~8 cm. In case of shorter cells, varying degrees of pulse overlap are expected. In contrast, none of these possibilities will be realized if the two pulses annihilate. In this case, only one discrete, undistorted AP should appear at each potential sensor. Which of these outcomes is realized in *Chara* internodes? To investigate this question, two APs were triggered simultaneously at both ends of the cell cylinder (Fig. 1D). In none of the experiments conducted (N=26 in n=10 cells; data for all control and collision experiments are publicly available [11]), an additional discrete AP or "M"-shaped pulse appeared at the extremal sensors. In one of the cells, one external sensor recorded a broadened AP. However, in a subsequent control experiment in the same cell a similar broadening was observed, which suggest that the reason for broadening is not a penetrated pulse. The lack of amplitude doubling in the central region of the cell (Fig. 2D and data in [11]), where collision occurs, provided further evidence that two APs do not superimpose and penetrate. Interestingly, the multisensor arrangement also allowed to extract the pulse propagation velocities for the control and collision experiments. In 19 of 25 measurements the velocity was increased (relative change on average ~10%) when two pulses were excited simultaneously. This essentially confirms an observation of Tasaki who reported an increase in conduction velocity when two nerve pulses approach each other head-on [4]. Taken together, these experiments with multiple sensors along the cell underline that two action potentials that meet head-on – within the conditions tested – neither summate and penetrate nor reflect off of each other.

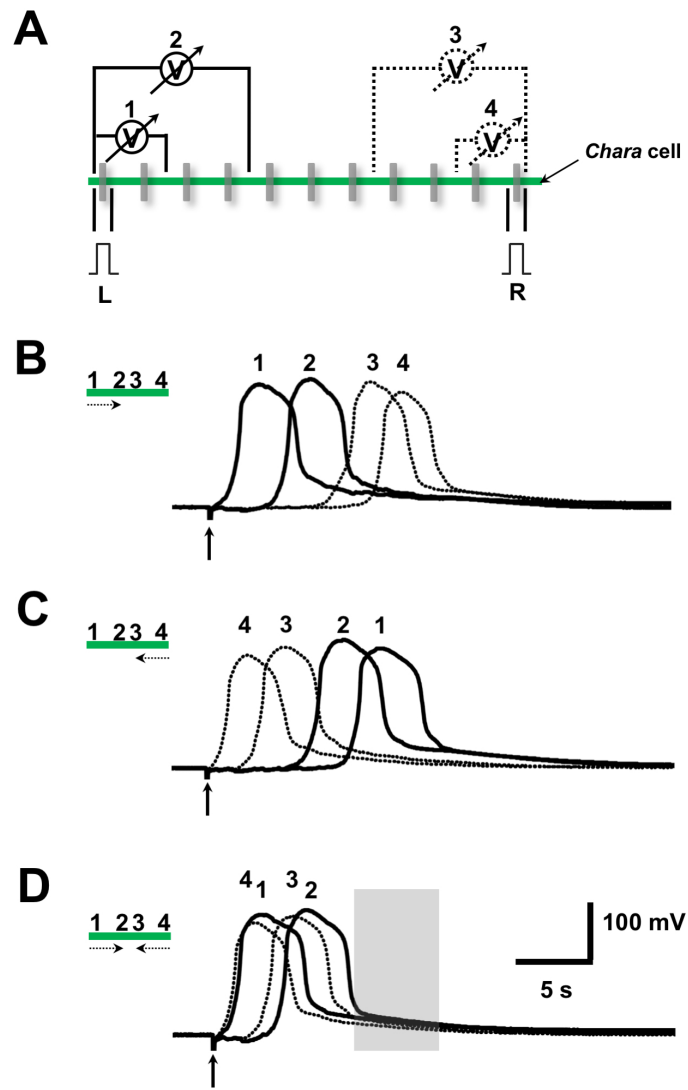

**Figure 2.** *Collision of two action potentials monitored with multiple sensors along the cell.* **(A)** The membrane potential was recorded at the two extremal regions (solid line (1) and dotted line (4)) and in the central parts (solid (2) and dotted (3)) of a *Chara* internode. **(B)** A single excitation at L led to the opposite propagation path as compared to a single excitation at R **(C)**. **(D)** Simultaneous excitation at L and R led to the head-on meeting of two APs in the central region of the cell. *Note*: At all sensors only one, undistorted AP was recorded which indicates that the pulses neither superimposed nor penetrated. Annihilation is most evident in the absence of additional pulses in the extremal sensors (grey region).

This conclusion is in agreement with previous findings in myelinated nerve fibers from frog [4], lobster axons [6], as well as ventral nerve cords of earthworms and the stomatogastric nervous system of the crab [5]. To the best of our knowledge, the only work which disagrees is the report of pulse penetration by Gonzalez-Perez et al. [7]. It will be valuable to address in future studies, if the latter findings can be replicated by other investigators[*]. Until then, the majority of results suggest that annihilation of two APs is a general characteristic in excitable plant cells as well as nerve fibers. Albeit penetration of two APs in *Chara* can be excluded, it is of interest whether annihilation of the two pulses is complete or partial. In 1 out of 26 collision experiments, an additional low-amplitude deflection of the membrane potential record was observed, which points to a residual pulse component (Fig. 1E). The existence of such a residual may be related to Tasaki's observation of record baseline fluctuations after collision of two APs in myelinated frog fibers [4]. Future studies could address if partial annihilation is generally a very rare event or if its likelihood increases for certain states of the excitable cell, for instance, when the refractory period becomes short (*e.g.* at elevated temperature).

---

[*] while the present work was under review, a debate of the study has taken place [36,37].

***3.3 Pulse collision and the physical nature of an action potential.*** At present, APs are considered to be electrical phenomena and their theoretical description is based on the formalism established by Hodgkin and Huxley [12]. The latter model, despite its widespread acceptance, has come under criticism. From a physical perspective, the severest objections are the existence of *reversible* mechanical, optical, thermal, etc. pulse components, which are not contained in a dissipative electrical framework. These discrepancies will not be discussed herein since they were laid out in detail by others (see *e.g.* [7,13–15]). The shortcomings of the classical electrical model have motivated the development of alternative theories of the action potential [7,13–15]. Kaufmann, for example, proposed that an AP is an acoustic[†] pulse in the lipid bilayer [14]. Recently, evidence in favor of this proposition has been obtained. In lipid monolayers, linear [16,17] as well as nonlinear waves [18] have been excited and detected by a variety of means (*e.g.* chemically, mechanically, electrically, etc.). The solitary waves in particular share several key characteristics with APs (threshold, all-or-none behavior, etc.) [18,19]. These pulses also propagate over extended distances without major changes to their shape. This behavior is usually attributed to particular nonlinear material properties which counteract dispersion of the wave packet. Solitary waves have been observed in many different fields such as hydrodynamics [20], optics [21], Bose-Einstein condensates [22] and during catalytic surface reactions [23].

Which predictions emerge from linear and nonlinear acoustics as pertaining to the head-on meeting of two pulses? In a recent work, it was suggested that the only conceivable outcome is superposition and penetration [5]. This is indeed expected in the realm of linear acoustics. In excitable cells, however, a linear relationship between stimulus and response is only observed in certain regimes. In postsynaptic membrane regions of neurons, for example, typical excitatory and inhibitory inputs summate [24]. This is reminiscent of wave superposition. The membrane of axons also responds linearly to hyperpolarizing stimuli as well as depolarizing stimuli with small amplitude [24] and such membrane potential displacements should superimpose. However, as the amplitude of depolarization of an axon is increased, the stimulus-response curve becomes nonlinear and an action potential is induced. Taken together, these observations demonstrate an important point, namely that linear as well as nonlinear perturbations can be excited in the same system. APs are phenomena in the nonlinear regime. There, the superposition principle does not apply and an extended spectrum of outcomes including complete and partial annihilation may be realized when two pulses meet. Some possibilities shall be discussed.

---

[†] It has to be emphasized that the term "acoustic" does not imply that an AP is a purely mechanical process. Rather, a propagating perturbation must manifest in local oscillations *of all thermodynamic observables* of the excitable medium (area density, charge density, internal energy density, etc.).

It has been suggested that APs are a class of nonlinear waves, so called solitons [7,15,25]. Upon meeting head-on, two solitons – similar to linear waves – penetrate and continue to propagate with widely preserved shape and speed [7,23,25]. This behavior is observed, for instance, for two such waves meeting on a water channel surface [20]. It was predicted that two solitons propagating in a lipid bilayer membrane should also penetrate [25]. The present results and the majority of reports in the literature [1,2,4–6] indicate, however, that pulse penetration is not observed in excitable membranes (note [7] as an exception). Considering this evidence, it is unlikely that APs are correctly described as solitons. It must be emphasized, however, that this does not falsify the proposition that an action potential is a nonlinear acoustic phenomenon. Other solitary waves annihilate upon meeting head-on, for example those in Bose Einstein condensates [22] as well as during catalytic surface reactions [23]. Interestingly, however, annihilation seems to be a comparatively rare scenario. One reason may be that annihilation requires a special type of nonlinearity in the system. There are indications, for example, that excitable cells reside close to a phase transition [13,26]. If the latter constitutes the nonlinearity that allows for the formation of an AP, the pulse could entail a phase change of the excitable medium. This transient realization of a different material phase may lead to pulse annihilation when two APs meet. In order to be able to penetrate, the pulses would have to propagate through this new phase. The latter, however, lacks the nonlinearity of the resting state and thus does not support this type of wave.

This leads us to the traditional explanation in the electrophysiological framework [5]. There, it is held firmly that two APs annihilate upon collision because of a refractory zone that trails the pulse [1–3,5]. In the electrical framework, the latter acquires its impenetrability from the properties of single molecules (*i.e.* ion channel proteins). Shortly after an AP, these proteins are believed to be in an inactive state [5]. Only after relaxation to their resting state, a subsequent excitation of the membrane will be possible. In this explanation, the excitable membrane is separated into active (ion channel proteins) and passive parts (lipid bilayer). This opens up a possibility to test the different theories of the AP. In acoustics, the propagation medium is not subdivided. The latter theory will be falsified if it is shown that the lipid bilayer remains in a constant thermodynamic state during an AP. This has often been presumed to be the case, because early measurements indicated that the electrical capacity of the excitable membrane does not change during activity [27]. Later experiments demonstrated, however, that this assumption was incorrect and that capacity shifts by ~25% during activity in squid axons [28]. Thus, it will be of interest to re-investigate this central issue by monitoring the thermodynamic state of an excitable membrane during an AP. In addition, it will be important to study if solitary waves in lipid monolayers annihilate upon collision. This will provide a direct test of the hypothesis that specific membrane proteins are required for pulse annihilation in a lipid membrane interface. Another question that remains is what happens at the site of collision. The simplest possibility is that the local temperature increases. This should, in principle, be observable with very sensitive thermal sensors [29]. However, in particular in biological materials, which have many degrees of freedom, it is conceivable that non-thermal transformations take place (budding/fusion of vesicles, changes of pH, transformation into chemical energy, etc.).

Finally, we would like to point out that there are several examples of other wave phenomena that annihilate upon meeting head-on. The latter was reported, for instance, in the Belousov-Zhabotinsky reaction system [30]. Waves of NADH and proton concentration which occur during glycolytic reactions in organelle-free yeast extracts also annihilate [31]. The same behavior was observed for intracellular waves in myocytes [32] and oocytes [33]. Further examples exist in multicellular systems. Mechanical stimulation of the nerve net of a sea pen is associated with luminescence waves and the latter extinguish upon running into each other [34]. Castro and Martins-Ferreira demonstrated in remarkable experiments with chick retina that this also holds true for two spreading depression waves in cortical tissue [35]. Taken together, these findings suggest *(i)* that specific molecules are unlikely to be the cause of annihilation because the phenomenon occurs in media with different composition and *(ii)* the intriguing possibility that different biological interfaces (plasma membrane, cytoplasm, cerebral cortical tissue, etc.) have been optimized for propagation of similar nonlinear waves.

## 4. Conclusions

When two action potentials were excited simultaneously at the ends of an excitable plant cell cylinder (*Chara* internode), they propagated towards each other and met head-on. The pulses annihilated in all experiments conducted. The majority of collision experiments carried out under standard conditions in nerve fibers came to the same conclusion. Thus, APs in excitable plant cells and axons share this as a common feature. It will be of interest to study if nonlinear solitary waves in lipid monolayers, which resemble action potentials, also extinguish upon collision. Further experiments should aim at understanding the mechanism that underlies annihilation. Concerning the latter, different theories make different predictions: While the relaxation of single molecules is crucial in the electrophysiological framework, an acoustic theory emphasizes the role of nonlinear material properties of the excitable medium.

## Acknowledgements

We thank K. Kaufmann for stimulating us to work on this problem and for numerous lectures and discussion sessions. This research did not receive any specific grant from funding agencies in the public, commercial, or not-for-profit sectors.